\documentclass[11pt]{article}
\usepackage{pst-all}
\usepackage{pstricks}
\usepackage{pstcol,pst-fill,pst-grad}
\usepackage{amsfonts}
\usepackage{graphicx}
\usepackage{amsmath}

\makeatletter \@addtoreset{equation}{section}

\newcommand{\be}{\begin{equation}}
\newcommand{\ee}{\end{equation}}
\newcommand{\bea}{\begin{eqnarray}}
\newcommand{\eea}{\end{eqnarray}}
\begin{document}
\date{}
\title{
\textbf{   Qubits  from Adinkra Graph Theory  via Colored  Toric Geometry }\\
\textbf{   } }
\author{ Y. Aadel$^{1}$,   A. Belhaj$^{2}$, Z. Benslimane$^{1}$,  M. B. Sedra$^{1}$, A. Segui$^{3}$
\hspace*{-8pt} \\
\\
{\small $^{1}$LHESIR,   D\'{e}partement de Physique, Facult\'{e} des
Sciences, Universit\'{e} Ibn Tofail }\\{ \small K\'{e}nitra,
Morocco} \\
{\small $^{2}$D\'epartement de Physique, Facult\'e
Polydisciplinaire, Universit\'e Sultan Moulay Slimane}\\{ \small
B\'eni Mellal, Morocco }
\\
 {\small $^{3}$Departamento de F\'isica Te\'orica, Universidad de Zaragoza,
E-50009-Zaragoza,
 Spain.}
 }  \maketitle

\begin{abstract}
We develop a new approach to deal with   qubit information systems
using   toric  geometry  and its  relation  to Adinkra graph theory.
More precisely, we link three different subjects namely toric
geometry, Adinkras and quantum information theory. This one to one
correspondence may be explored to attack  qubit system problems
using geometry considered as a powerful tool  to understand modern
physics including string theory. Concretely,  we examine  in some
details the cases of one, two, and three qubits, and we find that
they are associated with $\bf CP^1$, $ \bf CP^1\times CP^1$ and $\bf
CP^1\times CP^1\times CP^1$ toric varieties respectively. Using a
geometric procedure referred to as colored toric geometry, we show
that the qubit physics can be converted into a scenario handling
toric data of such manifolds by help of Adinkra graph theory.
Operations on toric information  can produce universal  quantum
gates.
\end{abstract}


\thispagestyle{empty}

\newpage \setcounter{page}{1} \newpage

\section{Introduction}
Toric geometry  is considered as a nice  tool to study complex
varieties used in physics including string  theory and related
models\cite{1,2}.  The key point  of this method is that the
geometric properties of such manifolds  are encoded in toric data
placed on a polytope consisting of vertices linked by edges.  The
vertices satisfy  toric constraint equations which have been
explored to solve many string theory problems such as  the absence
of non abelian gauge symmetries in  ten dimensional type II
superstring spectrums\cite{3}.

Moreover, toric geometry  has been also used to build mirror
manifolds providing  an  excellent  way to understand the extension
of T-duality in the presence of D-branes moving near  toric
 Calabi-Yau singularities using combinatorial calculations \cite{4}.  In
 particular,  these manifolds have been used in the context of    $N = 2 $  four dimensional quantum field theories
 in order to
obtain exact results using local mirror symmetry\cite{3}.  Besides
such applications, toric geometry has been also  explored  to
understand a class of  black hole solutions  obtained from  type II
superstrings on local Calabi-Yau manifolds \cite{5,6}.

Recently, the black hole physics has found a  place in quantum
information theory  using qubit  building blocks.  More precisely,
many connections  have been  established including the link with STU
black holes as proposed in \cite{7,8,9}.

More recently, an extension  to extremal black branes  derived from
the $T^n$ toroidal compactification of type IIA superstring have
been proposed in \cite{10}. Concretely, it  has been shown  that the
corresponding physics can be related to  $n$ qubit systems via the
real Hodge diagram  of such compact manifolds. The analysis has been
adopted to $T^{n|n}$ supermanifolds by supplementing fermionic
coordinates associated with the  superqubit formalism and its
relation to supersymmetric models.

The aim of this paper is to contribute to this program   by
introducing colored toric  geometry  and its relation to Adinkra
graph theory  to approach qubit information  systems.  An objective
here is to connect three different subjects namely toric geometry,
Adinkras and quantum information theory. This  link could be
explored to deal with  qubit systems using geometry considered as a
powerful tool  to understand modern physics. As an illustration, we
examine lower dimensional qubit systems. In particular, we consider
in some details the cases of one, two and three  qubits, we find
that they are linked with $\bf CP^1$, $\bf CP^1\times CP^1$ and $\bf
CP^1\times CP^1\times CP^1$ toric varieties respectively. Using a
geometric procedure referred to as colored toric geometry, we show
that the qubit physics can be converted into a scenario working
with toric data of such manifolds by help of Adinkra graph theory.

The present paper is organized as follows. Section 2 provides
materials  on how colored  toric geometry  may be used to discuss
qubit information systems. The connection with Adinkra graph theory
is investigated in Section 3 where focus is on an  one to one
correspondence between  Adinkras, colored toric geometry and qubit
systems. Operations on toric graphs  are employed in Section 4 when
studying universal  quantum gates. Section 5 is devoted to some
concluding remarks.

\section{      Colored
toric geometry  and Adinkras} Before giving  a colored toric
realization of qubit systems, we present an overview on  ordinary
toric geometry.  It has been realized that such a geometry is
considered as  a powerful tool to deal with complex Calabi-Yau
manifolds used in the  string theory compactification and related
subjects\cite{2}. Many examples have been elaborated in recent years
producing non trivial geometries.

 Roughly speaking, $n$-complex dimensional toric manifold, which we denote as
$M_{\triangle }^{n},$ is obtained  by considering the $(n+r)$%
-dimensional complex spaces $C^{n+r},$ parameterized by homogeneous
coordinates $\{x=(x_{1},x_{2},x_{3},...,x_{n+r})\},$ and $r$ toric
transformations  $T_{a}$  acting on the $x_{i}$'s as follows
\begin{equation}
T_{a}:x_{i}\rightarrow x_{i}\left( \lambda _{a}^{q_{i}^{a}}\right).
\end{equation}
Here,   $\lambda _{a}$'s are $r$ non vanishing  complex parameters.
For each $a$,  $q_{i}^{a}$  are integers which  called Mori vectors
encoding many geometrical information on the manifold and its
applications to   string theory  physics.
In fact,  these toric manifolds can be identified   with  the coset space $%
C^{n+r}/C^{\ast r}$.  In this way, the  nice  feature   is the toric
graphic realization.  Concretely, this realization is generally
represented by an integral polytope $\Delta $, namely a toric
diagram,   spanned by $(n+r)$ vertices ${v}_{i}$ of the standard
lattice $Z^{n}$. The toric data $\{v_i,q^a_i\}$ should satisfy the
following $ r$ relations
\begin{equation}
\sum_{i=0}^{n+r-1}q_{i}^{a}{v}_{i}=0,\qquad a=1,\ldots,r.
\end{equation}
Thus, these equations encode geometric data of $M_{\triangle }^{n}$.
In connection with lower dimensional field theory, it is worth
noting that  the $q_{i}^{a}$ integers  are interpreted, in the
${\cal N}=2$ gauged linear sigma model language, as the $U(1)^{r}$
gauge charges of ${\cal N}=2$ chiral multiples. Moreover, they have
also a nice geometric interpretation in terms of the intersections
of complex curves $C_{a}$ and divisors $D_i$ of $M_{\triangle}^{n}$
\cite{3,4,11}. This  remarkable link  has been explored in many
places in physics. In particular, it has been used  to build type
IIA local geometry.

The simplest example in toric geometry, playing  a  primordial  role
in the building block of higher dimensional toric varieties, is $\bf
CP^1$.  It is  defined by $ r = 1$ and the Mori  vector charge takes
the values $q_i = (1,1)$. This geometry has an U(1) toric action
$\bf CP^1$ acting as follows
\begin{equation}
z\to e^{i\theta} z,
\end{equation}
where $z=\frac{x_1}{x_2}$,   with    two fixed points $v_0$ and
$v_1$  placed on the real line. The latters  describing   the North
and south poles respectively of such a geometry, considered as  the
(real) two-sphere ${\bf S}^2\sim \bf CP^1$,  satisfy the following
constraint toric equation
\begin{equation}
v_0+v_1=0.
\end{equation}
In  toric geometry language, $\bf CP^1$  is represented by a  toric
graph identified with an interval $[v_0,v_1]$
  with
a circle on top.  The latter  vanishes  at the end points $v_0$ and
$v_1$.  This  toric  representation  can be easily  extended to the
$n$-dimensional   case using different ways. The natural one is the
projective space $\bf CP^n$. In this way, the  $\bf S^1$ circle
fibration, of $\bf CP^1$, will be replaced by ${\bf T}^n$ fibration
over an $n$-dimensional simplex (regular polytope). In fact, the
${\bf T}^n$ collapses to a ${\bf T}^{n-1}$ on each of the $n$ faces
of the simplex, and to a ${\bf T}^{n-2}$ on each of the
$(n-2)$-dimensional intersections of these faces, etc.

The second  way is to consider a class of  toric varieties  that we
are interested in here  given by a trivial product of one
dimensional projective spaces $\bf CP^1$'s admitting a similar
description.  We will show later on that this class  can be used to
elaborate a  graphic  representation of  quantum information systems
using ideas inspired by  Adinkra graph theory and related issues
[12-18]. For simplicity reason, we deal with  the case of $\bf CP^1
\times CP^1.$ For higher dimensional geometries $\bf
\bigotimes_{i=1}^n{CP}^{1}_i$ and their blow ups, the toric
descriptions  can be obtained   using a similar  way. In fact, they
are  $n$ dimensional toric manifolds exhibiting $U(1)^n$ toric
actions.  A close  inspection shows that there is a similarity
between toric graphs of such manifolds and qubit  systems using a
link with Adinkra graph theory. To make contact with quantum
systems, we reconsider the study of toric geometry by implementing a
new toric data associated with the color, producing a colored  toric
geometry. In this scenario, the toric data $\{v_i,q^a_i\}$ will be
replaced by
\begin{equation}
\{v_i, q^a_i, c_j,\;\;j=1,\ldots n \}.
\end{equation}
where   $c$ indicates the color of the edges linking the  vertices.
Roughly speaking,  the connection that we are after requires that
the toric graph should consist of $n+ r$ vertices and $n$ colors. In
fact, consider a special class of toric manifolds  associated with
$\bigotimes_{i=1}^n{CP}^{1}_i$ with $U(1)^n$ toric actions
exhibiting $2^n$ fixed points $v_i$. In toric geometry langauge, the
manifolds   are represented by $2^n$ vertices $v_i$ belonging to the
$Z^n$ lattice  satisfying $n$ toric equations. It is observed   that
these graphs share a strong resemblance with
 a particular class of  Adinkras  formed
by $2^n$ nodes connected  with
 $n$ colored edges  \cite{9}. These types of graphs  are called regular
 ones which can be used to present  graphically the $n$-qubit
 systems. At first sight,  the connection is not obvious.  However,
 our main  argument  is  based on the Betti    number  calculations.
 In fact,  these numbers  $b_i$  appear   in Adinkras and the
 corresponding toric graphs.  For $\bf CP^1 \times CP^1$, it is easy
 to calculate such numbers. They are given by
\begin{eqnarray}
b_0=1, \qquad b_2=2,\qquad
  b_4=1
\end{eqnarray}
Indeed, these numbers  can be identified with $(1,2,1)$  data used
in the  $n=2$  classification of Adinkras.

\section{ Andinkras and
colored toric geometry of qubits } Inspired   by   combinatorial
computations   in quantum physics, we explore colored   toric
geometry to deal with  qubit information systems  [19-29].
Concretely, we elaborate  a toric  description in terms  of a
 trivial fibration of one dimensional  projective space $\bf CP^1$'s. To
start, it is recalled that the  qubit  is a two state system which
can be realized, for instance, by    a  $1/2$ spin atom.
 The superposition state of  a single  qubit is generally  given by  the following
 Dirac notation
\begin{equation}
|\psi\rangle=a_0|0\rangle+a_1 |1\rangle
\end{equation}
 where $a_i$  are complex  numbers   satisfying the normalization
condition
\begin{equation}
|a_0|^2+|a_1 |^2=1.
\end{equation}
It is remarked  that this constraint  can be interpreted
geometrically in terms of the so called Bloch sphere, identified
with $ \frac{SU(2)}{U(1)}$ quotient  Lie  group \cite{1,2,3,4}. The
analysis can be extended to more than one qubit which has been used
to discuss entangled states. In fact, the two qubits are four
quantum level systems. Using the usual
 notation  $|i_1i_2\rangle=|i_1\rangle|i_2\rangle$,   the corresponding
 state superposition  can be expressed as follows
\begin{equation}
|\psi\rangle=\sum\limits_{i_1 i_2=0,1}a_{ i_1 i_2}|i_1
i_2\rangle=a_{00}|00\rangle+a_{10}
|10\rangle+a_{01}|01\rangle+a_{11} |11\rangle,
\end{equation}
where $a_{ij}$  are  complex numbers verifying   the normalization
condition
\begin{equation}
|a_{00}|^2+|a_{10}|^2+|a_{01}|^2+|a_{11}|^2=1
\end{equation}
describing the  $\bf CP^3$ projective space.  For $n$ qubits,  the
general state has the following form
\begin{equation}
\label{qudit} |\psi\rangle=\sum\limits_{i_1\ldots i_n=0,1}a_{
i_1\ldots i_n}|i_1 \ldots i_n\rangle,
\end{equation}
where $a_{ij}$  satisfy the normalization condition
\begin{equation}
\sum\limits_{i_1\ldots i_n=0,1}|a_{ i_1\ldots i_n}|^2=1.
\end{equation}
This condition defines   the  $\bf CP^{2^n-1}$ projective space
generalizing the Bloch sphere associated with $n=1$.\\ Roughly, the
qubit systems can be represented by colored toric diagrams having a
strong resemblance with
 a particular class of  bosonic  Adinkras,
 introduced in the study of supersymmetric
representation theory,   by  Gates  and its
group\cite{22,23,24,25,26,27,28}. In fact, there are many kinds of
such graphs. However,  we consider a particular class  called
regular one consisting  of $2^n$ vertices  linked by
 $n$ colored edges  as will be shown latter on.   An inspection,
 in  graph theory of Adinkras and toric varieties,    shows that we can propose the following
   correspondence connecting
three different subjects

\begin{table}[!ht]
\begin{center}
\begin{tabular}{|c|c|c|}
\hline  Adinkras &  Colored Toric Geometry &  Qubit systems
 \\ \hline  Vertices  & Fixed points (vertices)&  basis state \\ \hline
   Number of colors & Number of toric actions (Dimension) & Number
   of qubits
 \\ \hline
\end{tabular}
\end{center}
\label{tab2} \caption{This table presents an  one to one
correspondence between colored toric geometry, Adinkras and qubit
systems.}
\end{table}

To see how this works in practice, we   first change the  usual
toric geometry notation.  Inspired by combinatorial  formalism used
in quantum  information theory, the previous toric data can be
rewritten as follows
\begin{equation}
\sum\limits_{i_1\ldots i_n=0,1} q_{i_1\ldots i_n}^{a}{v}_{i_1\ldots
i_n}=0,\qquad a=1,\ldots,r,
\end{equation}
where the vertex subscripts indicate the corresponding  quantum
states. To illustrate this notation, we present a model associated
with  $\bf CP^1\times CP^1$ toric variety. This model is related to
$n=2$ Adinkras with (1,2,1) data as listed in  the classification.
In this case, the combinatorial Mori vectors can take the following
form
\begin{eqnarray}
q^1_{i_1i_2}&=&(q^1_{00},q^1_{01},q^1_{10},q^1_{11})= (1,0,0,1)\\
\nonumber q^2_{i_1i_2}&=&(q^2_{00},q^2_{01},q^2_{10},q^2_{11})=
(0,1,1,0).
\end{eqnarray}
The  manifold corresponds to the toric equations
\begin{equation}
\sum\limits_{i_1i_2=0,1} q_{i_1 i_2}^{a}{v}_{i_1 i_2}=0,\qquad
a=1,2.
\end{equation}

 In colored  toric geometry langauge,
it is represented by $4$ vertices ${v}_{i_1 i_2}$, belonging to
$Z^2$, linked by   four edges with  two different colors $c_1$ and
$c_2$.  The toric data require the following vertices
 \begin{equation}
 v_{00}=(-1,0),\; v_{01}=(0,1),
\;v_{10}=(0,-1), \;v_{11}=(1,0)
\end{equation}
with two colors. These data  can be encoded in  a toric graph
describing  two qubits and it  is illustrated in figure 1.
\begin{figure}[h]
        \begin{center}
            \includegraphics[scale=0.5]{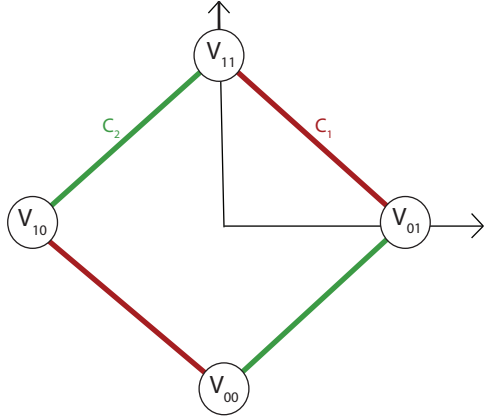}
        \end{center}
        \caption{Toric Adinkra graph representation of  for $ n=2 $ qubits.}
\end{figure}
\\

\section{ Quantum  gates from geometry}
Having examined the qubit object, we move now  to  build  the
quantum gates using colored toric geometry and Adinkra graph theory.
The general study is beyond of the scope of this paper. We consider,
however, lower dimensional cases.  To do so, it is recalled that the
classical gates can be obtained by combining Boolean operations as
AND, OR,XOR, NOT and NAND.  In fact, these operations act on input
classical bits, taking  two values 0 and 1, to produce new bits as
output results. In quantum computation,  gates are unitary operators
in a $2^n$ dimensional Hilbert space. In connection with
representation theory, they can be represented by $2^n\times 2^n$
matrix, belonging to $SU(2^n)$ Lie group, satisfying the following
properties
\begin{eqnarray}
U^+=U^{-1},\qquad  det\;U=1
\end{eqnarray}
As  in the classical case, there is an universal notation for the
gates depending on the input qubit number. The latters are
considered as building blocks  for constructing  circuits and
transistors.  For 1-qubit computation, the  usual one is  called NOT
acting  on the basis state as follows
\begin{eqnarray}
|i_1\rangle \to | \overline{i_1}\rangle
\end{eqnarray}
In this  toric geometry language, this operation corresponds  to
permuting the  two toric vertices of $\bf CP^1$
\begin{eqnarray}
\sigma: v_0  \leftrightarrow  v_1
\end{eqnarray}
This operation  can be represented by  the following matrix
 \begin{eqnarray}
 \left(
              \begin{array}{cc}
                0 & 1 \\
                1 & 0 \\
              \end{array}
            \right)
\end{eqnarray}
which can be identified  with $U_{NOT}$ defining the NOT quantum
gate. In this case,  it  is worth noting that the corresponding
color operation is
trivial since we have only one.\\
  For 2-qubits, there are many
universal gates. As mentioned previously,  this  system is
associated with the toric geometry of $\bf CP^1\times CP^1.$ Unlike
the 1-qubit case corresponding to $\bf CP^1\times CP^1$ , the
quantum  systems involve two different data namely the vertices and
colors. Based on this observation, such data will produce two kinds
of operations:
\begin{enumerate}
  \item color actions
  \item vertex actions.
\end{enumerate}
In fact, these  operations can  produce
 CNOT and SWAP gates.  To get such gates,   we fix the color action
 according to Adinkra orders used in the corresponding notation.  Following   the colored toric realization
 of the 2-qubits, the color actions can be  formulated as follows
\begin{eqnarray}
c_1 &:& |i_1i_2\rangle \to |\overline{i_1}i_2\rangle\\\nonumber c_2
&:& |i_1i_2\rangle \to |i_1\overline{i_2}\rangle.
\end{eqnarray}
In this color language,    the  CNOT gate
\begin{eqnarray}
CNOT=\left(
       \begin{array}{cccc}
         1 & 0 & 0 & 0 \\
         0 & 1 & 0 & 0 \\
         0 & 0 & 0 & 1 \\
         0 & 0 & 1 & 0\\
       \end{array}
     \right)
\end{eqnarray}
can be obtained by using the following actions
\begin{eqnarray}
c_1 &\to& c _1 \\ \nonumber c_2 &\to& c_2\otimes c_1.
\end{eqnarray}
A close inspection shows  that the SWAP gate
\begin{eqnarray}
 SWAP=\left(
       \begin{array}{cccc}
         1 & 0 & 0 & 0 \\
         0 & 0  & 1 & 0 \\
         0 & 1 & 0 & 0 \\
         0& 0 & 0 & 1 \\
       \end{array}
     \right)
\end{eqnarray}
can be derived  from the following permutation  action
\begin{eqnarray}
c_1 \to c _2.
\end{eqnarray}
We expect that  this   analysis can be adopted to higher dimensional
toric manifolds.  For simplicity reason, we consider the geometry
associated with  TOFFOLI gate being an universal gate acting on a
3-qubit.  It is remarked  that geometry can be identified with the
blow up of $\bf CP^1\times CP^1\times CP^1$ toric manifold. In
colored toric geometry language, this manifold is described  by the
following equations
\begin{equation}
\sum\limits_{i_1i_2i_3=0,1} q_{i_1i_2 i_3}^{a}{v}_{i_1i_2
i_3}=0,\qquad a=1,\ldots,5,
\end{equation}
where $2^3$ vertices $v_{i_1i_2i_3}$ belong to $Z^3$. They are
connected by three different colors $c_1$, $c_2$ and $c_3$. These
combinatorial equations   can be solved by the following Mori
vectors
\begin{eqnarray}
q_{i_1i_2i_3}^1&=&(1,0,0,1,0,0,0,0)\nonumber\\
q_{i_1i_2i_3}^2&=&(0,1,0,0,1,0,0,0)\nonumber\\
q_{i_1i_2i_3}^3&=&(0,0,1,0,0,1,0,0)\\
q_{i_1i_2i_3}^4&=&(1,-1,0,0,0,0,1,0)\nonumber\\
q_{i_1i_2i_3}^5&=&(0,0,1,1,0,0,0,1).  \nonumber
\end{eqnarray}
Thus, the   corresponding vertices $v_{i_1i_2i_3}$ are given by
\begin{eqnarray}
v_{000}=(1,0,0),\; v_{100}=(0,1,0), \;v_{010}=(0,0,1),
\;v_{001}=(-1,0,0)\\
\nonumber
v_{110}=(0,-1,0),\;v_{101}=(0,0,-1),\;v_{110}=(-1,1,0),\;v_{111}=(0,-1,-1),
\end{eqnarray}
and they   are connected with three colors. This  representation can
be illustrated in figure 2.
\begin{figure}[h]
        \begin{center}
            \includegraphics[scale=0.5]{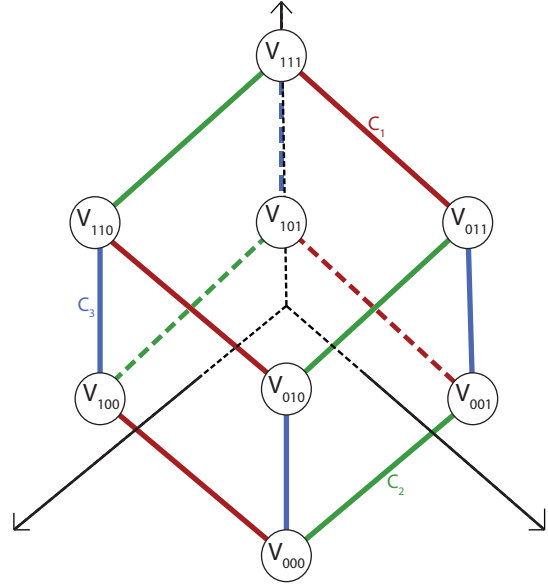}
        \end{center}
        \caption{Regular Adinkra graphic  representation  for $n=3$.}
\end{figure}

The   TOFFOLI  gate  represented by
$2^3\times 2^3$ matrix
\begin{eqnarray}TOFFOLI=
\left(
  \begin{array}{cccccccc}
    1& 0 & 0 & 0 & 0 & 0 & 0 & 0 \\
    0 & 1 & 0 & 0 & 0 & 0 & 0 & 0 \\
    0 & 0 & 1 & 0 & 0 & 0 & 0 & 0 \\
    0 & 0 & 0 & 1 & 0 & 0 & 0 & 0 \\
    0& 0 & 0 & 0 & 1 & 0 & 0 & 0 \\
    0& 0 & 0 & 0 & 0 & 1 & 0 & 0 \\
    0 & 0 & 0 & 0 & 0 & 0 & 0 & 1 \\
    0 & 0 & 0 & 0 & 0 & 0 & 1 & 0 \\
  \end{array}
\right)
\end{eqnarray}
can be obtained by the following color transformation
\begin{eqnarray}
c_1 &\to& c _1 \nonumber \\  c_2 &\to& c_2\\  c_3 &\to& c_3\otimes
c_2\otimes c_1.\nonumber
\end{eqnarray}
We expect  that this analysis  can be pushed further to deal  with
other toric varieties having non trivial  Betti numbers.
\section{Conclusion}

Using   toric  geometry/ Adinkras correspondence, we  have discussed
qubit systems. More precisely, we have presented an  one to one
correspondence between  three different subjects namely toric
geometry, Adinkras and quantum information theory.  We believe that
this work  may  be explored to attack  qubit system problems using
geometry considered as a powerful tool to understand  modern
physics. In particular, we have considered in some details the cases
of one, two and three qubits, and we find that they are associated
with $\bf CP^1$, $\bf CP^1\times CP^1$ and $\bf CP^1\times
CP^1\times CP^1$ toric varieties respectively. Developing  a
geometric procedure referred to as colored toric geometry, we have
revealed that the qubit physics can be converted into a scenario
turning  toric data of such manifolds by help of Adinkra graph
theory. We have shown  that operations on such data can produce
universal quantum gates.

This work comes up with many open questions. A natural one is to
examine super-projective  spaces. We expect that this issue can be
related to superqubit systems. Another question is to investigate
the entanglement states in the context of toric geometry and its
application including mirror symmetry. Instead of giving a
speculation, we prefer  to comeback these open questions   in
future.

 \vspace{1cm}

{\bf Acknowledgments}: AB would like to thank the Departamento de
F\'{\i}sica Te\'{o}rica, Universidad de Zaragoza for kind
hospitality. He  would like to thank also  Diaz Family, Montanez
Family and Segui Family for kind hospitality in Spain. The authors
thank F. Falceto for interesting discussions. AS is supported by
FPA2012-35453.

\end{document}